\documentstyle[psfig,preprint,prl,aps]{revtex}

\begin{document}

\draft

\title{Collective Effects in the Horizontal Transport of Vertically
Vibrated Granular Layers}

\author{P. Tegzes and T. Vicsek}

\address{Department of Atomic Physics, E\"otv\"os University,
Budapest, Puskin u 5-7, 1088 Hungary}

\author{D.C. Rapaport}

\address{Department of Physics, Bar-Ilan University, Ramat-Gan 52900, Israel}

\date{\today}

\maketitle

\begin{abstract}

Motivated by recent advances in the investigation of fluctuation-driven
ratchets and flows in excited granular media, we have carried out
experimental and simulational studies to explore the horizontal
transport of granular particles in a vertically vibrated system whose
base has a sawtooth-shaped profile.  The resulting material flow
exhibits novel collective behavior, both as a function of the number of
layers of particles and the driving frequency; in particular, under
certain conditions, increasing the layer thickness leads to a {\it
reversal of the current}, while the onset of transport as a function of
frequency can occur either {\it gradually or suddenly} in a manner
reminiscent of a phase transition.

\end{abstract}


\narrowtext

The best known and most common transport mechanisms involve gradients of
external fields or chemical potentials that extend over the distance
traveled by the moving objects.  However, recent theoretical studies
have shown that  there are processes in far from equilibrium systems
possessing vectorial symmetry that can bias thermal noise type
fluctuations and  induce macroscopic motion on the basis of purely
local effects.  This mechanism is expected to be essential for the
operation of molecular combustion motors responsible for many kinds of
biological motion; it has also been demonstrated experimentally in
simple physical systems \cite{RoSaAj94,Fau95}, indicating that it could
lead to new technological developments such as nanoscale devices or
novel types of particle separators.  Motivated by both of these
possibilities, as well as by interesting new results for flows in
excited granular materials
\cite{JaNaBe96,DoFaLa89,ClDuRa92,EhJaKa95,PaBe193,KnJaNa93}, we have
carried out a series of experimental and simulational studies that
explore the manner in which granular particles are {\em horizontally}
transported by means of {\em vertical} vibration.

In the corresponding theoretical models -- known as ``thermal ratchets''
--  fluctuation-driven transport phenomena can be interpreted in
terms of overdamped Brownian particles moving through a periodic but
asymmetric, one-dimensional potential in the presence
of nonequilibrium fluctuations \cite{AjPr92,Mag93,AsBi94,DoHoRi94}.
Typically, a sawtooth-shaped potential is considered, and the nonlinear
fluctuations are represented either by additional random forces or by
switching between two different potentials.  Collective effects
occurring during the fluctuation-driven motion have also been considered
\cite{DeAj96,DeVi95,JuPr95}, leading to a number of unusual effects that
include current reversal as a function of particle density.

The investigation of an analogous transport mechanism for granular
materials is an appealing idea, both conceptually and practically. By
carrying out experiments -- both real and numerical -- on granular
materials  vibrated vertically by a base with a sawtooth profile, it
is possible to achieve a fascinating combination of two topics of
considerable current interest -- ratchets and granular flows. A number of
recent papers have focused on vibration-driven granular flow, and the
details of the resulting convection patterns have been examined, both by
direct observation \cite{DoFaLa89,ClDuRa92,PaBe193,KnJaNa93} and by
magnetic resonance imaging \cite{EhJaKa95,NaAlCa93}. Granular convection
has also been simulated numerically by several groups; the study most
closely related to the present work deals with the horizontal transport
that occurs when the base is forced to vibrate in an asymmetric manner
\cite{GaHeSo92}.

The present paper describes an investigation of the horizontal flow of
granular material confined between two upright concentric cylinders
undergoing vertical vibration. In order to induce transport, the height of
the annular base between the cylinders has a periodic, piecewise-linear
profile (in other words, it is sawtooth-like). We observe novel collective
behavior in the resulting material flow, both as functions of the number
of particle layers and the driving frequency. The most conspicuous
features, for the experimental parameters used here, are that increasing
the layer thickness results in a {\em reversal} of the current, and that
the onset of transport as a function of frequency takes place in a manner
analogous to a phase transition with an exponent $\beta \simeq 0.5$.
Numerical simulations supporting these experimental findings are briefly
described; these results also suggest the possibility of {\it further
non-trivial transitions} as the parameters governing the system are
varied.

Figure 1 shows a schematic view of the experimental apparatus.  To
achieve a quasi-two-dimensional system without boundaries in the
direction of the expected flow the granular material is placed between
two concentric glass cylinders \cite{PaBe193}.  The mean diameter of the
cylinders is 10\,cm, while the gap between the cylinders is either 3 or
5\,mm.  A ring filling the gap between the cylinders, with a sawtooth
profile on its upper surface, is mounted on the base of the container.
This ring is made of one of the following materials: PVC (soft),
aluminum,
 or danamid
(a hard, elastically responding synthetic material), 
and different sawtooth shapes are used.  The entire assembly
is vertically vibrated with a displacement that depends sinusoidally on
time.

Two types of granular media are used in the experiments, monodisperse
glass balls and quasi-ellipsoidal plastic beads (see the inset of Fig.~2).
The glass balls are nearly spherical with diameter 3.3 mm $\pm$ 2\%. The
plastic beads have a much greater size dispersion: two of the axes are
approximately equal in length and lie in the range 2.4--3.0\,mm, while the
third axis is 1.2--1.7\,mm. As shown in the inset of Fig.~2, the size of
each sawtooth is similar to that of the particles.

Provided the frequency is sufficiently large, the vertical vibration
causes horizontal flow of the entire granular layer. This bulk motion is
reproducible over repeated experiments. The average flow velocity is
determined by tracking individual tracer particles visible through the
transparent cylinder walls. In order to average out fluctuations, the
particles are allowed to travel large distances; depending on the size of
the fluctuations this distance is between 1.5 and 6\,m (equal to 5--20
times the circumference of the system). Each point shown in the graphs is
an average over 3--6 tracer particles.

Figure 2 shows the horizontal flow velocity as a function of the number
of particles for various possible systems.  The actual sawtooth and
particle shapes are also indicated.  Positive velocities are defined to
be in the direction for which the left-hand edge of the sawtooth has the
steeper slope (from left to the right for these cases).  The vibration
amplitude and frequency are $A = 2$\,mm and $f = 25$\,Hz; the
dimensionless acceleration $\Gamma = (2 \pi f)^2 A / g$ is an important
quantity for vibrated granular systems, so that here $\Gamma = 5$.

We have observed a variety of qualitatively different kinds of behavior,
some of which can be interpreted by simple geometrical arguments.  The
most surprising phenomenon is that in certain cases the velocity changes
sign; in other words, the flow {\em direction} depends on the layer
thickness.  In some cases the curves are monotonically decreasing, while
others have well defined maxima.  Altering the particle shape reduces
the velocity and shifts the location of the maximum, but the shape of
the curve remains unchanged.  The only feature common
to the different curves is that beyond a certain layer thickness the
velocity magnitude decreases as further layers are added.

According to our studies of a simplified geometrical model the following
qualitative argument can be used to explain the observed current
reversal: There is an intermediate size and asymmetry of the teeth for
which a single ball falling from a range of near-vertical angles bounces
back to the left (negative direction) in most of the cases.  This effect
is enhanced by rotation, due to friction between the ball and the tooth.
However, if there are many particles present, this mechanism is
destroyed, and on average, the direction of the motion of particles will
become positive (the "natural" direction for this geometry); this
corresponds to the usual ratcheting mechanism characterized by larger
distances traveled by the particles along the smaller slope with
occasional jumps over to the next valley between the teeth
\cite{AjPr92,Mag93,AsBi94,DoHoRi94,Bug87}.  There is no net current for
symmetric teeth, although the motion of the particles is very
interesting in that situation as well \cite{Duran}.

We interpret the existence of maxima as follows: If only a few
particles are are present, their motion is erratic, with large jumps
in random directions.  As the number of particles is increased, due to a
inelastic collapse-like process, the particles start to move coherently,
and this more ordered motion, together with the right
frequency, seems to give rise to a kind of resonant behavior as far as
transport is concerned.  Similar maxima can also be observed in
molecular motor calculations and simulations.  Thick layers move
more slowly because of inelastic damping.

Figure 3 shows the $\Gamma$ dependence of the flow velocity for a system
of 200 balls (amounting to 4 layers) for constant $A$.
 Flow occurs only
above a critical acceleration $\Gamma_c \simeq 1.7$. Above this critical
value the velocity appears to follow a power law
\begin{equation}
 v(\Gamma) \propto (\Gamma - \Gamma_c)^{0.48},
\end{equation}
suggesting that the onset of flow resembles the kind of phase transition
observed in hydrodynamic instabilities such as thermal convection
\cite{BeDu84}.


In an attempt to learn about the the general nature of this class of
systems, we have carried out a series of two-dimensional simulations using
a molecular (or granular) dynamics approach \cite{Bar94,Her95,Rap95}. The
grains are modeled as relatively hard, rotating disks that undergo damped
collisions; details of the general method and the interparticle forces
appear elsewhere \cite{HiRaRa97} (the damping constants here are
$\gamma_s = \gamma_n = 5$), and this same system (with a flat base and
without rotation) has been used in a recent study of surface excitations
\cite{Rap98}. The base is constructed of disks similar to those
representing the grains, but only 1/3 the size, positioned (with some
overlap) to produce the required sawtooth profile; all the disks forming
the base oscillate vertically with the appropriate amplitude and
frequency. The lateral boundaries are periodic.

The simulations reveal very complex behavior that is not only in
qualitative agreement with what is observed experimentally, but also
suggests further directions for laboratory exploration.  Horizontal
transport is clearly present, but the magnitude and direction of the
flow depend -- often in a complex fashion -- on the details of the
sawtooth profile (tooth width, height and degree of asymmetry), on $A$
and $f$, and also on the number of particle layers and the damping
constants.  Given the difficulty in constructing a model that embodies
all of the actual collisional properties of the experimental system, our
goal is presently limited to the qualitative reproduction of experiment.

Space permits just a single demonstration of the complex behavior.
Fig.~4 shows the mean horizontal velocity ($v$) as a function of the
number of particle layers ($nl$).  The three curves are for systems that
differ in sawtooth details, with all other parameters the same.  In
terms of dimensionless MD units (in which particle diameters are of
order unity) the system is of width 90, $A = 1$ and $\Gamma
= 2$; other details appear in the caption.  Each data point is an
average over 500 base oscillation cycles (the first 100 cycles of each
run are excluded to allow the decay of initial transients).  The results
show the same kinds of behavior apparent in experiment, namely, positive
or negative velocities over a range of layer depths, or a transition
from negative to positive values.

The remarkable result that emerges from both experiment and simulation is
that the flow direction can change as the layer thickness varies. This is
entirely unexpected and requires further investigation; the only related
behavior of which we are aware is the alternating current direction in a
model of collectively moving interacting Brownian particles in a
``flashing'' \cite{DeAj96} ratchet potential. Furthermore, the experiment
described above shows a relatively sharp onset of horizontal motion, a
result also obtained by simulation (not shown) for certain parameter
values; in other cases, simulation reveals a more gradual onset of motion
with subsequent oscillatory frequency dependence, something not yet
observed in the laboratory.


In conclusion, we have investigated granular transport in a system
inspired by models of molecular motors and have observed, both
experimentally and numerically, that the behavior depends in a complex
manner on the parameters characterizing the system. These results ought to
stimulate further research into this fascinating class of problems.

Acknowledgments: Useful discussions with I. Derenyi are acknowledged.
This work was supported in part by FKFP Grant No.\ 0203/1977 and OTKA
Grant No.\ T019299.  One of us (DCR) acknowledges the support of the
Israel Science Foundation.


\begin{figure}
\vfill
\psfig{figure=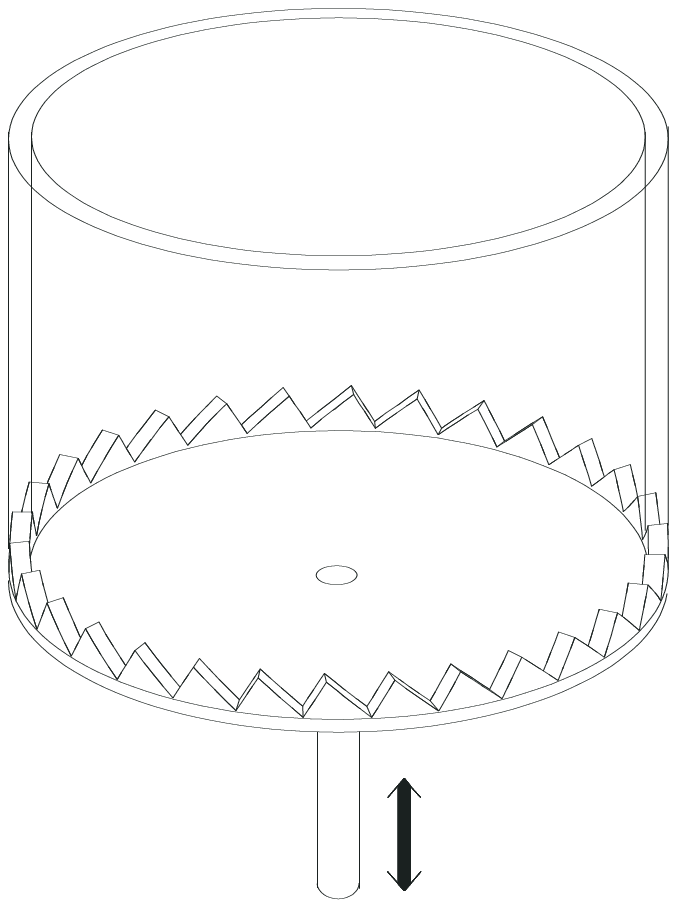,height=15cm,width=15cm}
\vfill
\caption{Diagram of the experimental apparatus. The granular material is placed between the two glass cylinders and the whole assembly is subjected to sinusoidal vertical vibration.}
\end{figure}
\vfill
\eject

\begin{figure}
\psfig{figure=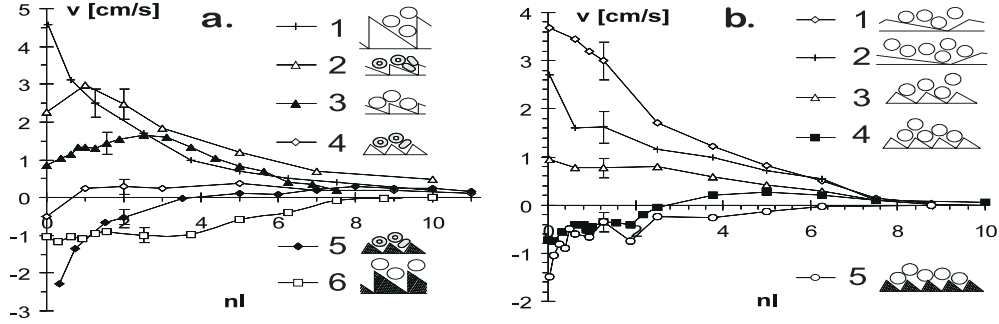,width=15cm,height=15cm}
\caption{Horizontal velocity as a function of the number of
particles. The various curves represent measurements for various sawtooth shapes (given by  horizontal projections  $w_1$ and $w_2$ of the left and right parts of a tooth and its height $h$), materials and two kinds of
particles. {\it a:}
(1) (0\,mm,10\,mm,10\,mm) PVC sawtooth and glass balls;
(2) (0\,mm,6\,mm,3\,mm) PVC sawtooth and plastic beads;
(3) (0\,mm,6\,mm,3\,mm) PVC sawtooth and glass balls;
(4) (1\,mm,3\,mm,3\,mm) PVC sawtooth and plastic beads;
(5) (1\,mm,3\,mm,3\,mm) hard plastic sawtooth and plastic beads;
(6) (0\,mm,7\,mm,7\,mm) hard plastic sawtooth and glass balls;
{\it b:}
All of the curves correspond to glass balls. The parameters of the
sawteeth were
(1) (4\,mm,12\,mm,3\,mm) PVC;
(2) (6\,mm,18\,mm,3\,mm) PVC;
(3) (1.5\,mm,4.5\,mm,3\,mm) PVC;
(4) (1.25\,mm,3.75\,mm,3\,mm) PVC;
(5) (1\,mm,3\,mm,3\,mm) hard plastic;
The amplitude and
frequency are $A = 2$\,mm, $f = 25$\,Hz.}
\end{figure}
\eject

\begin{figure} 
\vfill
\psfig{figure=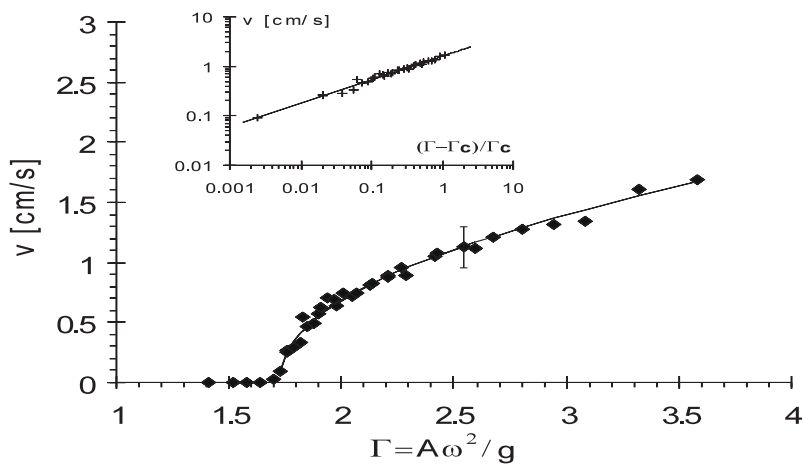,width=15cm,height=15cm}
\vfill
\caption{Horizontal velocity $v$ as a function of
the
dimensionless acceleration $\Gamma$ at constant amplitude ($A =
2$\,mm).
The experiment is for a strongly asymmetric aluminium sawtooth
($w_1=0\,mm$, $w_2=12\,mm$, $h=7\,mm$ and 200 glass
balls.  In the inset we display the data on a log-log scale for $v$
close to the transition as a function of
$(\Gamma-\Gamma_c)/\Gamma_c$ where
$\Gamma_c=1.7$. The slope of the fitted line is 0.48. } 
\end{figure}
\vfill
\eject

\begin{figure}
\vfill
\psfig{figure=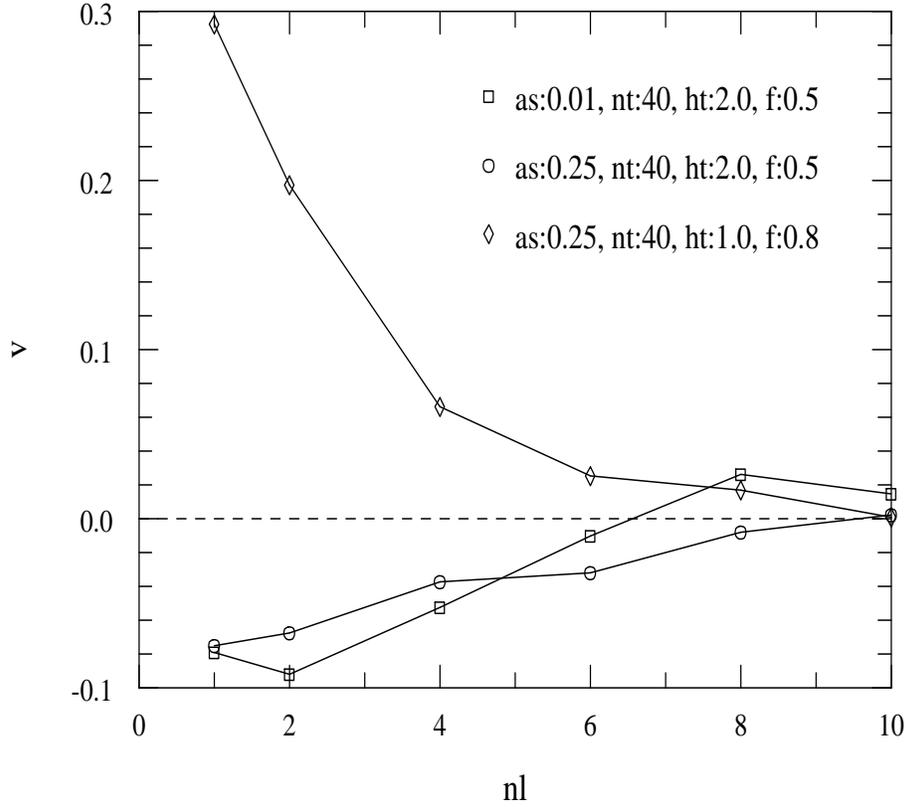,width=15cm,height=15cm}
\vfill
\caption{
Simulated horizontal velocity ($v$ -- in dimensionless units) as a
function of the number of particle layers ($nl$); ''as'' denotes sawtooth
asymmetry (0.01 corresponds to an almost vertical left edge, 0.25 to edges
with horizontal components having a 1:3 ratio), ''nt'' is the number of
sawteeth in the base, ''ht'' is the sawtooth height, ''f'' is the
dimensionless frequency; typical spread
of the velocity results, based on repeated runs with different initial
conditions, is $\pm 0.01$.}
\end{figure}
\vfill

\end{document}